\documentclass[%
 reprint,
 superscriptaddress,
 amsmath,amssymb,
 aps,
]{revtex4-2}

\usepackage{graphicx}
\usepackage{dcolumn}
\usepackage{bm}
\usepackage{color}
\DeclareMathAlphabet\mathbfcal{OMS}{cmsy}{b}{n}

\begin{document}

\preprint{APS/123-QED}

\title{Terahertz Radiation from Laser-Ionized Plasmas}

\author{Kathryn A. Wolfinger}
\email{kathryn.wolfinger@colorado.edu.}
\author{Gregory R. Werner}%
\affiliation{%
Center for Integrated Plasma Studies, Physics Department, University of Colorado, 390 UCB, Boulder, CO 80309, USA
}%

\author{John R. Cary}
\affiliation{%
Center for Integrated Plasma Studies, Physics Department, University of Colorado, 390 UCB, Boulder, CO 80309, USA
}%
\affiliation{%
Tech-X Corp., 5621 Arapahoe Avenue, Suite A, Boulder, Colorado 80303, USA
}%

\date{\today}

\begin{abstract}

Numerical simulations of laser-plasma interactions demonstrate the generation of axially polarized electromagnetic pulses (EMPs) that radiate away energy with a characteristic frequency determined by the plasma frequency, in the THz range for typical laser wakefield acceleration experiments. This is confirmed by full 2D electromagnetic particle-in-cell simulations, as well as by a ponderomotively-driven reduced model that captures the EMP generation essentials. When the laser’s pulse length matches the plasma wavelength, that pulse’s fractional energy losses to the wakefield are independent of the pulse width, while the losses to the EMP are inversely dependent on the pulse width.

\end{abstract}

\maketitle

The interaction of an intense laser pulse with a gas target ionizes that gas to produce a column of plasma via field ionization~\cite{Amm86}. As this laser pulse propagates through the newly created plasma, its ponderomotive potential drives an electrostatic wakefield within the plasma, as first observed in 1993 \cite{Ham93, Ham94}. Essentially, the ponderomotive potential pushes the lighter electrons farther than the significantly heavier ions, establishing a charge separation that drives the Langmuir-wave wakefield. Sufficiently strong wakefields will establish axial fields that can accelerate electrons up to several GeV over centimeter-scales, providing energies comparable to those of larger and more expensive conventional accelerators. The details of this process, called laser wakefield acceleration (LWFA), have been reviewed in \cite{Esa09}. In addition to these electrostatic wakefield effects, previous simulations in 2D have revealed that the laser's ponderomotive potential also drives electromagnetic pulses (EMPs) that radiate from the plasma \cite{Spr04}. We investigate these EMPs in 2D full electromagnetic particle-in-cell (PIC) simulations, as well as in reduced fluid models that capture the essentials of the laser-plasma interaction with significantly less computational effort. 

In particular, in this paper we investigate these electromagnetic effects to determine the amount of energy the EMPs carry off from the laser pulse in comparison with the wakefield's energy. Extracting the EMP effects from the PIC noise with a moving-average technique and comparing those results with our reduced fluid model, we have also found that these EMPs have characteristic frequencies which correspond to the plasma frequencies, and therefore the densities, of the plasmas they radiate from, with values in the terahertz (THz) range. There is substantial interest in the theory behind THz generation as a radiation source for areas like spectroscopy \cite{Gaa06}, medical applications \cite{Pic06}, and single-shot imaging \cite{Jia98, Dob06}. So far several methods have been used to generate THz emission, including optical rectification in the presence of plasma \cite{Kre04, Xie06}, transition-Cherenkov radiation \cite{DAm07}, counter-propagating lasers \cite{Hur19}, and ponderomotive acceleration \cite{Spr04, Tho07}.

\emph{The simulation setup} is as follows. We model this laser-plasma interaction with the VSim/Vorpal application \cite{Nie04,VSi21} in 2D-Cartesian geometry, with longitudinal ($x$) and one transverse ($y$) dimension. The electromagnetic PIC simulations are initialized with a room temperature, uniform, neutral hydrogen gas of density $n_0$. Two different densities are used: $n_0=2.4\times 10^{18}\:\textrm{cm}^{-3}$ is chosen from the density range of recent LWFA experiments, and $n_0=6\times 10^{16}\:\textrm{cm}^{-3}$ is chosen to explore the EMP behavior at a lower density. Time step sizes are determined to satisfy the Courant stability limit, set at $c\Delta t/\Delta x = 0.7$ for both densities considered. The prescribed laser pulse propagates in the $+x$ direction, and the simulation grid shifts in $+x$ at the speed of light, $c$, to move along with the pulse. It is a circularly polarized laser pulse of intensity $I_0 = 1.3\times 10^{16}\:\textrm{W}/\:\textrm{cm}^{2}$ with an $800\:\textrm{nm}$ wavelength, a $40\:\mu$m radial FWHM in $y$, and a $50\:\textrm{fs}$ ($15\:\mu$m) pulse length.
\begin{figure}[ht]
\centering
\includegraphics[width=8.6cm]{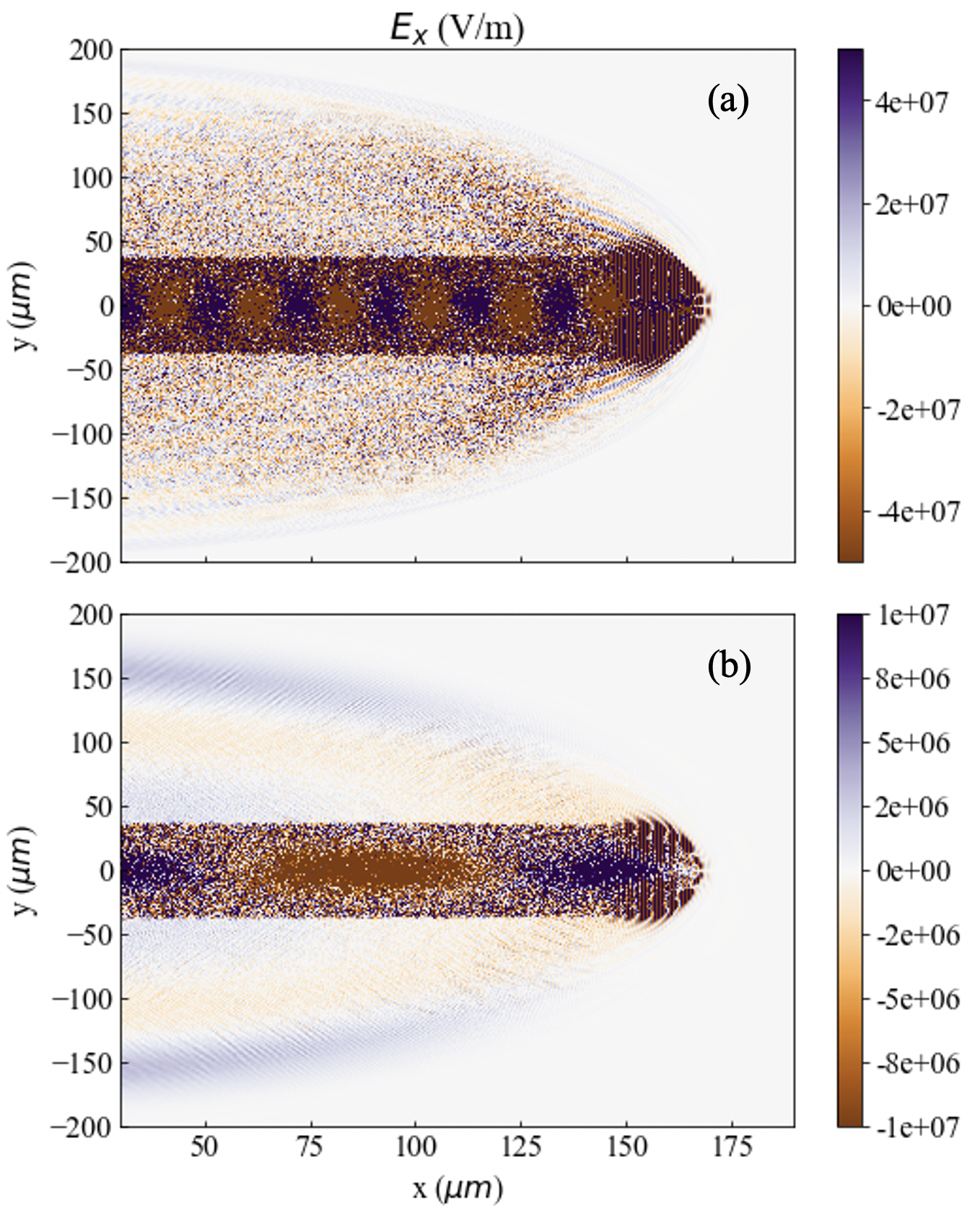}
\caption{PIC results showing the axial electric field, $E_x$, created by a laser propagating in the $+x$ direction through hydrogen gas of densities (a) $n_0=2.4\times 10^{18}\:\textrm{cm}^{-3}$ and (b) $n_0=6\times 10^{16}\:\textrm{cm}^{-3}$, ionizing plasmas. The stronger field within the plasma can be seen at $|y|\lesssim 35\:\mu$m. The wakefield's longitudinal oscillations of $E_x$ within the plasma columns, visible along $y = 0$, correspond to the plasma wavelengths for their respective densities.}
\label{fig:MovingWindow}
\end{figure}

As the pulse propagates through the hydrogen gas, it produces plasma of density $n$ via field ionization, with those ionization rates taken from the Ammosov-Delone-Krainov (ADK) model \cite{Amm86}. For this model, the rate depends on the laser field amplitude $E_L$ at the particle positions and the ionization threshold $U_i$ of the neutral gas: 
\begin{eqnarray} \label{eq:ADKrate}
\text{Rate}_{\rm ADK} =&& ~\left(4.13\times 10^{16}\:\textrm{s}^{-1}\right) \frac{Z^2}{4\pi n_{\rm eff}^3} \left( \frac{2 E_h Z^3}{E_L n_{\rm eff}^3}\right)^{2n_{\rm eff}-1}\nonumber\\
&&\times\left( \frac{2}{n_{\rm eff}} \right) ^{2n_{\rm eff}}\text{exp}\left( \frac{-2E_h Z^3}{3E_L n_{\rm eff}^3} +2n_{\rm eff}\right)
\end{eqnarray}
where $Z$ is the degree of ionization, $E_h = 5\times 10^{11}\:\textrm{V/m}$, and $n_{\rm eff}=Z/\sqrt{U_i / 13.6~\text{eV}}$ is the effective principal quantum number. For hydrogen gas we use $U_i=15.4\:\textrm{eV}$, and not $13.6\:\textrm{eV}$, since it is appropriate for molecular hydrogen by taking into account the presence of other processes, like dissociation and excitation \cite{Shi93,Sha19}. Once the plasma is formed by fully ionizing the gas, the laser interacts with it to generate both a wakefield and an EMP.
\begin{figure}[ht]
\centering
\includegraphics[width=8.6cm]{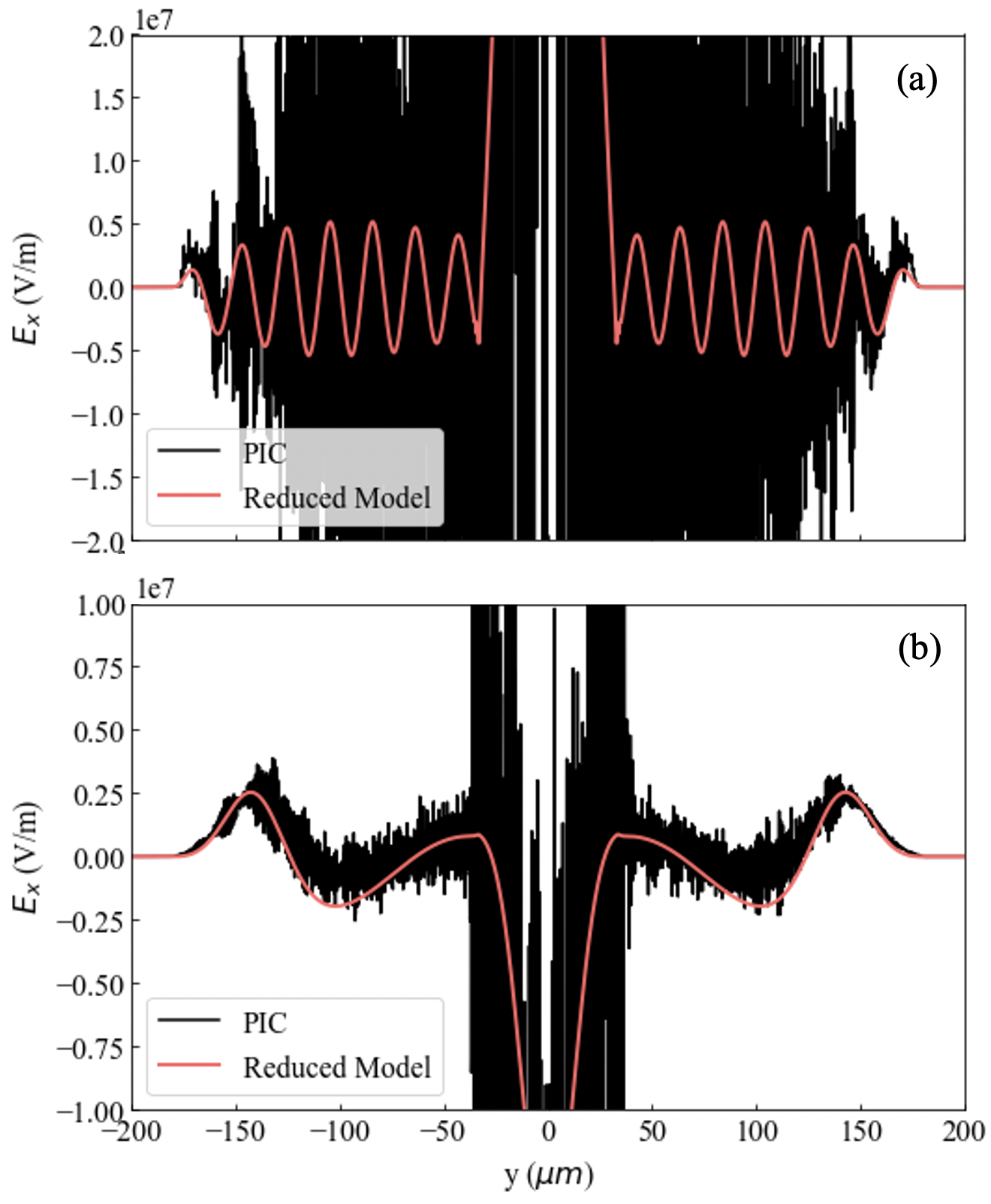}
\caption{The electric field, $E_x$, from Fig. \ref{fig:MovingWindow} plotted vs $y$ for $x=53\mu m$. The PIC results are overlaid with $E_x$ vs $y$ plots from a reduced fluid model for the EMP generation, as seen in Fig.~\ref{fig:SimpleModel}. The high density case (a) of $n_0=2.4\times 10^{18}\:\textrm{cm}^{-3}$ shows no clear structure due to overwhelming PIC noise, and the lower density case (b) of $n_0=6\times 10^{16}\:\textrm{cm}^{-3}$ shows a simple pulse EMP which matches with the results of the reduced fluid model.}
\label{fig:HistoryAvg}
\end{figure}

\emph{The simulation results} can be seen in Fig. \ref{fig:MovingWindow}, which demonstrates the EMP and the wakefield values of $E_x$  for two of these PIC simulations. The plasma slabs ionized by the laser can be seen at $|y|\lesssim 35\:\mu$m, extending behind the laser pulse in x. For the wakefields, the longitudinal oscillations of $E_x$ within the plasma are visible along $y = 0$. The lengths of these wake oscillations in $x$ correspond to the ``plasma wavelength,'' $\lambda_p = 2\pi c / \omega_p$, for their respective densities, where $\omega_p$ is the plasma frequency. Specifically, for case (a) in Fig. \ref{fig:MovingWindow}, the longitudinal oscillations of $E_x$ within the plasma have a wavelength of approximately $22\:\mu$m in $x$. This length corresponds to a plasma frequency of $8.6\times 10^{13}\:\textrm{s}^{-1}$ and a density of $n=2.3\times 10^{18}\:\textrm{cm}^{-3}$, which is close to the simulation's neutral gas density. For case (b) in Fig. \ref{fig:MovingWindow}, the wakefield oscillations are approximately $135\:\mu$m in $x$. This length corresponds to a plasma frequency of $1.4\times 10^{13}\:\textrm{s}^{-1}$ and a density of $n=6.1\times 10^{16}\:\textrm{cm}^{-3}$, which also approximates that simulation's neutral gas density. The EMPs can be seen propagating away from the plasma slabs on either side in $y$, polarized in the direction of laser propagation. At the higher density in (a), the EMP is only slightly visible amidst PIC noise. In the lower density case of (b), a simple pulse structure can been seen propagating away from the plasma. Simulations with linearly polarized laser pulses yield essentially indistinguishable EMPs for the same laser intensity. In Fig. \ref{fig:HistoryAvg} we plot $E_x$ vs $y$, taken from approximately $x=53\:\mu m$ in Fig. \ref{fig:MovingWindow}, to show the peak field values of the EMP. For the higher density of $n_0=2.4\times 10^{18}\:\textrm{cm}^{-3}$, the inherent PIC noise drastically overshadows the EMP .

The essential physics of the EMP and wake behavior can be captured by a 2D reduced model that includes transverse electromagnetic waves in an unmagnetized, inhomogeneous cold fluid plasma. These waves are driven by the electric current from the laser pulse's ponderomotive force. Unlike the previous PIC simulations and the fluid model of \cite{Spr04}, this simple model is designed to isolate the critical EMP-generating effects. It uses prescribed laser fields, instead of a self-consistent laser like in \cite{Spr04}, and neglects both collisions and density variations after the initial ionization by the laser pulse. Even without these effects, it captures the essentials of the wakefield and the EMP generation, while also doing so for significantly less computational effort and with less noise. Specifically, we take the ansatz that the laser's magnetic and electric fields are of the form
\begin{equation}\label{eq:EMFields}
    \bm{E_\perp} = \bm{E}_{\rm max} \sin\left(\omega t\right)\exp \left\{-\frac{1}{2}\left[\left(\frac{y}{\sigma_w}\right)^2+\left(\frac{\frac{x}{c}-t}{\sigma_t}\right)^2\right] \right\}
\end{equation}
with a sinusoid of frequency $\omega = 2\pi c /\lambda$ multiplied by a slowly varying envelope, $\sigma_t$, such that $\omega \sigma_t \gg 1$, where $\sigma_t\equiv [2\sqrt{2 \ln 2}]^{-1} \textrm{FWHM}_\textrm{time}$ and $\sigma_w\equiv [2\sqrt{2 \ln 2}]^{-1} \textrm{FWHM}_\textrm{width}$. From this, the ponderomotive force can be calculated as
\begin{equation}
    \bm{F}_{\rm pond} = - \frac{e^2}{4 m \omega^2}\bm{\nabla}\langle E_\perp^2\rangle
\end{equation}
and included in the equation of motion as
\begin{equation}\label{forceAxial}
    \frac{\partial}{\partial t} \bm{v} = \frac{e}{m} \bm{E} - \frac{e^2}{4 m^2 \omega^2} \bm{\nabla} \langle E_\perp^2\rangle
\end{equation}
where $\langle E_\perp^2\rangle$ is the time-average of the squared fields over the fast sinusoidal oscillations. Taking the time derivative of Ampere’s law, combined with the current density $\bm{j}=ne\bm{v}$, we get 
\begin{equation}\label{derivAmp}
    \frac{\partial^2}{\partial t^2} \bm{E} = -\frac{ne}{\epsilon_0} \frac{\partial}{\partial t} \bm{v}+c^2 \frac{\partial}{\partial t} \left(\bm{\nabla}\times\bm{B}\right).
\end{equation}
By substituting Eq. \ref{forceAxial} into Eq. \ref{derivAmp}, and using Faraday's law, we arrive at two coupled equations for the electric field in a 2D-Cartesian model, where $\partial/\partial z = 0$: 
\begin{subequations}
\label{eq:reducedModel}
\begin{equation}
\frac{\partial^2}{\partial t^2} E_x + \omega_p^2 E_x + c^2 \left(\frac{\partial^2 E_y}{\partial y \partial x}-\frac{\partial^2 E_x}{\partial^2 y}\right) = \frac{e \omega_p^2}{4 m \omega^2} \frac{\partial}{\partial x} \langle E_\perp^2\rangle\label{subeq:1}
\end{equation}
\begin{equation}
\frac{\partial^2}{\partial t^2} E_y + \omega_p^2 E_y + c^2 \left(\frac{\partial^2 E_x}{\partial x \partial y}-\frac{\partial^2 E_y}{\partial^2 x}\right) = \frac{e \omega_p^2}{4 m \omega^2} \frac{\partial}{\partial y} \langle E_\perp^2\rangle,\label{subeq:2}
\end{equation}
\end{subequations}
where the plasma density $n$ is included in Eq. \ref{eq:reducedModel} through the plasma frequency $\omega_p=(n e^2/m \epsilon_0)^{1/2}$ and is calculated using the ADK model (Eq. \ref{eq:ADKrate}) as
\begin{equation}\label{eq:density}
    \frac{\partial}{\partial t}n\left(x,y,t\right) = \left(n_0-n\right)*{\rm rate}_{\rm ADK}.
\end{equation}
We develop a simple finite-difference PDE solver on a grid that moves with the laser pulse to obtain the solutions of Eqs. \ref{eq:reducedModel} and \ref{eq:density}.

\begin{figure}[ht]
\centering
\includegraphics[width=8.6cm]{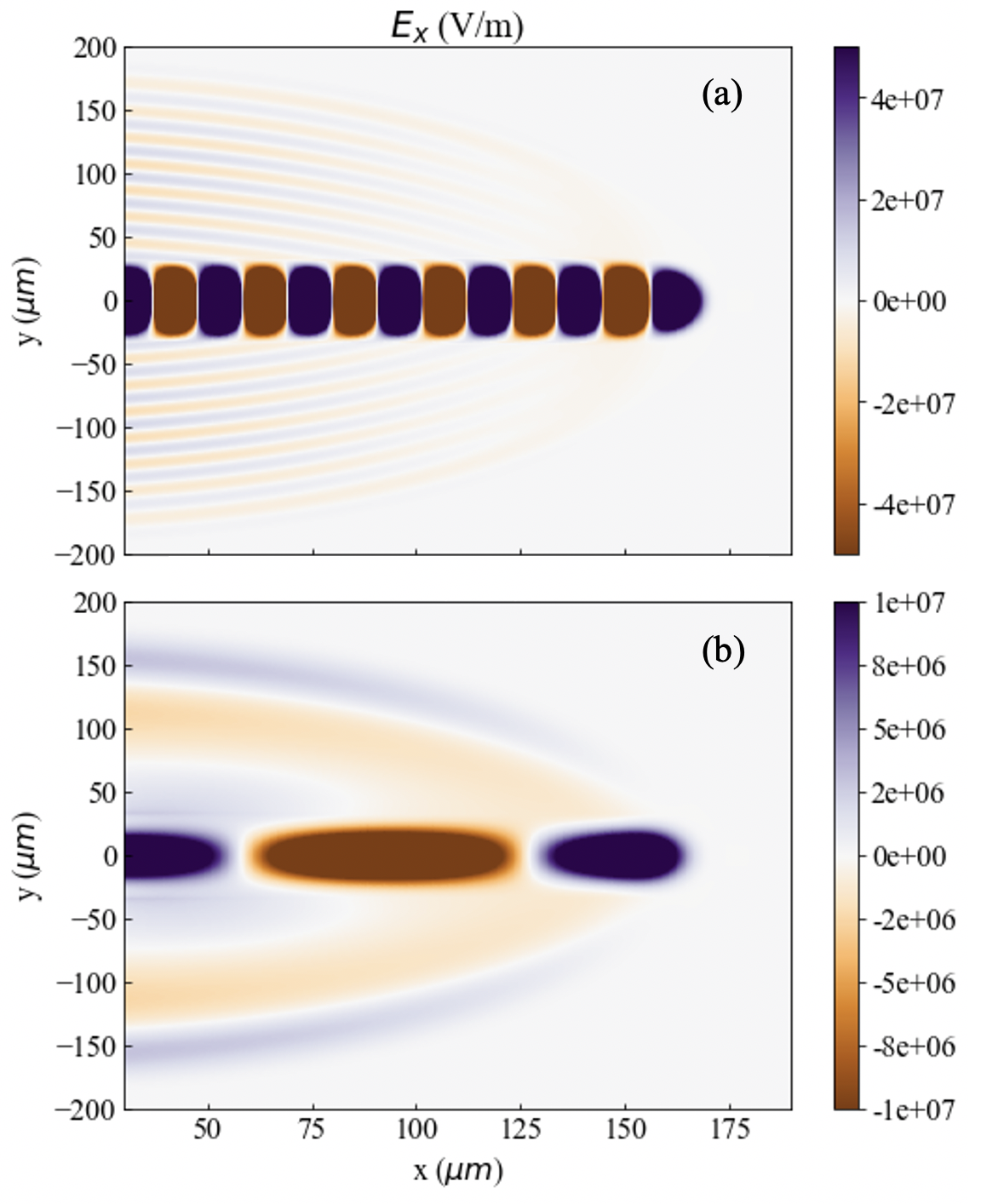}
\caption{Example plots from simulating Eq. \ref{eq:reducedModel} to be compared with Fig. \ref{fig:MovingWindow} for (a) $n_0=2.4\times 10^{18}\:\textrm{cm}^{-3}$ and (b) $n_0=6\times 10^{16}\:\textrm{cm}^{-3}$. The reduced model has less noise than the PIC results, especially for the higher density case (a).}
\label{fig:SimpleModel}
\end{figure}

\begin{figure}[ht]
\centering
\includegraphics[width=8.6cm]{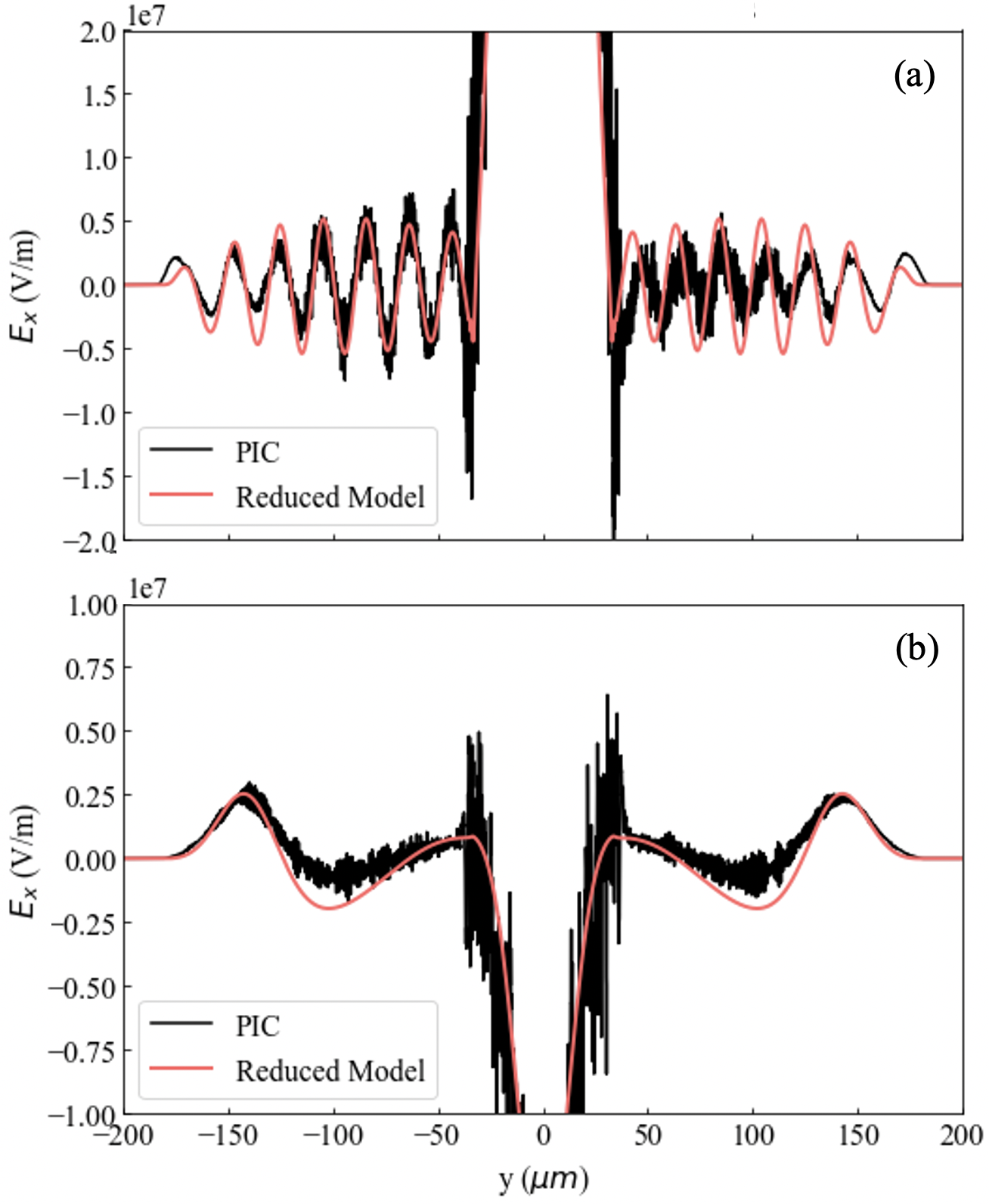}
\caption{$E_x$ vs $y$ from Fig. \ref{fig:MovingWindow}, produced by sampling the field at a set distance behind the laser pulse and averaging it over time. (a) The time-averaged lineout for the high density case of $n_0=2.4\times 10^{18}\:\textrm{cm}^{-3}$ shows a ringing EMP propagating from the plasma center, which was not visible in Fig. \ref{fig:MovingWindow} or Fig. \ref{fig:HistoryAvg} due to overwhelming PIC noise. (b) The lineout for the lower density of $n_0=6\times 10^{16}\:\textrm{cm}^{-3}$ also shows the simple-pulse EMP with more clarity. In both cases, the PIC lineouts agree with the overlaid reduced model lineout from Fig. \ref{fig:SimpleModel}.}
\label{fig:SMLineouts}
\end{figure}
\begin{figure}[ht]
\centering
\includegraphics[width=8.6cm]{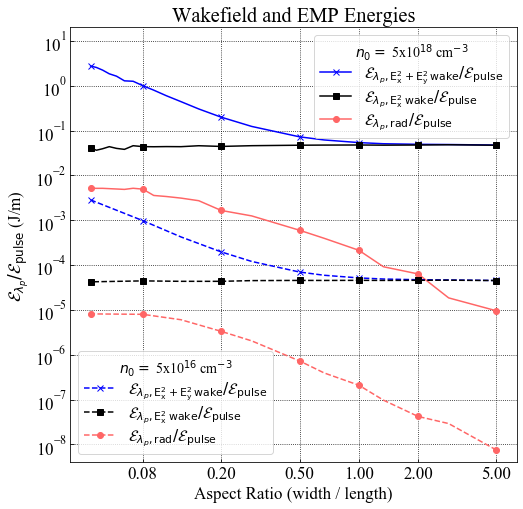}
\caption{Calculated the fractional energy loss of the drive laser pulse to the wakefield and the EMP, extracted from the reduced model simulations of Eq. \ref{eq:reducedModel}, for two densities ($n_0=5\times 10^{16}\:\textrm{cm}^{-3}$ and $n_0=5\times 10^{18}\:\textrm{cm}^{-3}$).}
\label{fig:Energies}
\end{figure}

The results of this reduced model, as shown in Fig. \ref{fig:SimpleModel}, can be compared with the PIC simulation results in Fig. \ref{fig:MovingWindow} and demonstrate that we can replicate both the EMP’s ringing nature in Fig. \ref{fig:MovingWindow}a and its simple-pulse structure in Fig. \ref{fig:MovingWindow}b by considering a ponderomotive driver. In addition, we replicate the plasma wavelengths and the pulses' characteristic frequencies, without the significant noise seen in the PIC simulations. We can further compare the reduced model with the PIC results by viewing the plots of $E_x$ vs $y$: in Fig. \ref{fig:HistoryAvg} for $n_0=6\times 10^{16}\:\textrm{cm}^{-3}$ there is agreement between PIC and the reduced model, but for $n_0=2.4\times 10^{18}\:\textrm{cm}^{-3}$ the EMP was initially overshadowed by noise inherent to the PIC simulations. To overcome that noise, we average the PIC results over time by recording the $E_x$ field at a set distance behind the laser pulse (and not at a set value in $x$), as seen in Fig. \ref{fig:SMLineouts}. With this averaging, there is good agreement between PIC and the reduced model, validating the plasma response to the ponderomotive driver as the source of the EMP, independent of plasma density fluctuations. 

These two simulations, with the same drive pulse length and different densities, generate EMPs with very different characteristic frequencies that match the background plasma frequencies. Specifically, the EMPs' characteristic frequency spectrums peak at approximately $9.4\times 10^{13}\:\textrm{s}^{-1}$ and $1.6\times 10^{13}\:\textrm{s}^{-1}$ for the high and low density cases, respectively. This contradicts the claim in \cite{Spr04} that the EMP's characteristic frequency matches the pulse length. However, it agrees with the claim that the EMP is not caused by local density variations: since the same EMP behavior is seen in both the PIC and the reduced model simulations, even though the reduced model does not include density fluctuations, the EMP cannot be caused by local variations of the plasma density from the background, or $n_0$, plasma density. In addition to these two cases, shown in Fig. \ref{fig:MovingWindow}-\ref{fig:SMLineouts}, we created a PIC simulation where the plasma wavelength ($150 \mu m$) was an order or magnitude larger than the pulse length ($15 \mu m$) and chose the pulse width ($75 \mu m$) such that the EMP oscillated repeatedly, or ``rang.'' For this PIC simulation, replicated with the reduced model, the EMP's characteristic frequency spectrum peaked at approximately $1.4\times 10^{13}\:\textrm{s}^{-1}$, matching the plasma frequency.

To further demonstrate the reduced model's usefulness, we examine the dependence of EMP strength on the pulse shape. Specifically, to explore the EMP and wakefield energies in the reduced model, we keep $E_{\rm max}=\sqrt{2I_0/c\epsilon_0}$ and now match the pulse to the plasma wavelength, $\sigma_t = [2\sqrt{2 \ln 2}c]^{-1}\lambda_p$, ideal for exciting the wakefields used for LWFA. This is done for two different densities: $n_0=5\times 10^{16}\:\textrm{cm}^{-3}$ and $n_0=5\times 10^{18}\:\textrm{cm}^{-3}$. Then, we vary $\sigma_w$ according to an aspect ratio defined as
\begin{equation}\label{eq:aspect_ratio}
    {\rm Aspect\: Ratio} = \frac{\sigma_w}{c\sigma_t}
\end{equation}
and ranging from 0.04 to 5.0. As the laser pulse propagates, it loses energy to the electrostatic wakefields (and associated particle motion), which are essentially stationary with negligible group velocity, as well as to the outward-radiating EMP fields. Considering a box around the plasma column of length $\lambda_p$ in~$x$, extending from $-y'$ to $+y'$ where $y'>y_p$ (and arbitrary length $L_z$ in the unsimulated direction), we estimate both the wakefield energy left behind and the EMP field energy radiated outward as the pulse travels through the box.  The total energy radiated through both the outer walls is the integrated Poynting flux:
\begin{eqnarray}\label{eq:emp_energy}
    \mathcal{E}_{\lambda_p,\rm rad} = 2 L_z \int_0^\infty dt \, 
    \int_{\lambda_p} dx\, \frac{-1}{\mu_0}E_x(x,y',t) B_{z}(x,y',t).
\end{eqnarray}
The wakefield energy in $E_x$ in the box is:
\begin{subequations}\label{eq:emp_energy}
\begin{equation}
    \mathcal{E}_{\lambda_p,E_x^2, \rm wake} = L_z \int_{\lambda_p}dx\int_{-y_p}^{y_p} dy \frac{\epsilon_0}{2} E_x^2,\label{subeq:1}
\end{equation}
\begin{equation}
    \mathcal{E}_{\lambda_p,E_x^2+E_y^2, \rm wake} = L_z \int_{\lambda_p}dx\int_{-y_p}^{y_p} dy \frac{\epsilon_0}{2} (E_x^2+E_y^2).\label{subeq:2}
\end{equation}
\end{subequations}

For this comparison, we leave out particle kinetic energy, which scales proportional to the field energy. We also consider the wakefield energy in $E_y$, which becomes significant only for thin pulses, $\sigma_w \ll c\sigma_t$. In Fig.~\ref{fig:Energies}, we graph these energies normalized by the laser pulse energy,
\begin{equation}\label{eq:pulse_energy}
    \mathcal{E}_{\rm pulse} = L_z \times 2\pi\sigma_w\sigma_t I
;\end{equation}
thus, $\mathcal{E}_{\lambda_p,\rm rad}/\mathcal{E}_{\rm pulse}$ and $\mathcal{E}_{\lambda_p,\rm wake}/\mathcal{E}_{\rm pulse}$ yield the fractions of the pulse field energy lost to the EMP and the wakefield over one plasma wavelength.

We see that, when considering only the energy in $E_x$, $\mathcal{E}_{\lambda_p,\rm wake}/\mathcal{E}_{\rm pulse}$ is independent of the aspect ratio because $\mathcal{E}_{\lambda_p,\rm wake} \propto \sigma_w\sigma_t I$, and the pulse intensity $I$ is kept constant.  If we include the energy in the $E_y$ component, this is still the case for large aspect ratios, where the wakefield can be treated one-dimensionally.  For thinner pulses, however, $y$-gradients become significant and the ponderomotive force excites motion in $y$, yielding substantial $E_y$ fields. Compared with the wakefield values, $\mathcal{E}_{\lambda_p,\rm rad}/\mathcal{E}_{\rm pulse}$ decreases with increasing aspect ratio as $\mathcal{E}_{\lambda_p,\rm rad}/\mathcal{E}_{\rm pulse} \sim \sigma_w^{-1/2}$ for larger aspect ratios and saturates for small aspect ratios. Finally, the energies in Fig. \ref{fig:Energies} vary with density like $n_0^{3/2}$. This factor arises from the density dependence of the wakefield amplitude in linear ($a \ll 1$) plasma waves, $E_x\propto\sqrt{n_0}$, combined with the normalization of the resulting energies by the laser pulse energy \cite{Esa09}.

\emph{In conclusion}, we have shown that in 2D PIC simulations the laser-plasma interaction results in the radiation of a radially propagating, axially polarized electromagnetic pulse, for which the EMP's oscillation frequency depends on the generated plasma's density. We have further developed a reduced model that replicates the results of the full PIC simulations, reproducing the wakefield and EMP with only essential physics. This simplified model is easily generalized to 2D $r$-$z$ cylindrical coordinates for a similar computational expense. It also yields the same EMP as the PIC simulations without the presence of the noise seen in them and for significantly less computational effort. This allows for the consideration of the drive laser pulse's energy that is lost to the wakefield and EMP. When the laser pulse's length is matched to the plasma wavelength, the normalized EMP and wakefield energies vary with density as $n^{3/2}$. The wakefield energy is largely independent of the pulse width when normalized by the laser energy, while the EMP energy varies inversely with the laser's aspect ratio. Varying the pulse length from the plasma wavelength in future numerical studies will clarify the use of this EMP as a diagnostic for plasma radius and density.

\begin{acknowledgments}
This material is based upon work supported by the National Science Foundation under grant no. 1734281. This research used resources of the National Energy Research Scientific Computing Center, a DOE Office of Science User Facility supported by the Office of Science of the U.S. Department of Energy under Contract No. DE-AC02-05CH11231.
\end{acknowledgments}

\nocite{*}

\bibliography{Bibliography}

\end{document}